\documentclass[aps,prl,twocolumn,epsf,epsfig,amsmath]{revtex4}

\usepackage{latexsym}
\usepackage{hyperref}
\usepackage{times}
\usepackage{graphicx}
\usepackage{amsmath}
\usepackage{multirow}
\usepackage{amsthm}
\usepackage{dcolumn}
\usepackage{bm}
\usepackage{ textcomp }
\usepackage{color}
\usepackage{amssymb}
\usepackage{xcolor}
\usepackage{easyReview}
\usepackage{mathdots}
\usepackage{float}
\usepackage{lineno}
\usepackage{todonotes}
\usepackage{array}
\usepackage{orcidlink}

\newcommand{\beq}{\begin{eqnarray}}
\newcommand{\eeq}{\end{eqnarray}}

\begin{document}
\title{Proof that Momentum Mixing Hatsugai Kohmoto equals the Twisted Hubbard Model }
\author{Yuting Bai$^{1}$,\orcidlink{0009-0007-1551-3932}}
\email{yutingb2@illinois.edu}   

\author{Philip W. Phillips$^{1}$}
\email{dimer@illinois.edu}

\affiliation{$^1$Department of Physics and Institute of Condensed Matter Theory, University of Illinois at Urbana-Champaign, Urbana, IL 61801, USA}

\begin{abstract}

We prove formally that the momentum-mixing Hatsugai-Kohmoto model (MMHK) is the Hubbard model with a twist.  With this result in tow, we rely on the proof of Watanabe's that  two models which differ by a twist must have the same bulk physics.  Consequently, we have proven that MMHK=Hubbard in the charge sector.  

\end{abstract}
\date{December 2024}


\maketitle

The dichotomy between localized and itinerant electrons  leads to two distinct kinds of electronic eigenstates in quantum materials.  For example, 
while the band picture of metals places electrons in Bloch states with a well-defined momentum, Mott physics is thought to reside in the opposite real-space or local limit where electrons live on atomic sites.  Enshrining the latter is the Hubbard model in which an on-site repulsion frustrates the motion of electrons throughout the lattice.  Because of the hopping term, no eigenstate of the Hubbard model preserves on-site particle number.  As a result, understanding the Mott gap in real space is elusive. In momentum space, the problem is equally hopeless as the on-site repulsion mixes all momenta.  Nonetheless, numerics\cite{DMFT,ZhengScience2017,Huangnpj2018,QinPRX2020,shiwei1,shiwei2,XuPRL2013,DengPRL2013,ParkPRL2008,KancharlaPRB2008,maier,Maipnas2022,MaiNC2023,WernerPRB2009,RohringerRMP2018,LeBlancPRX2015,SCHAFER2016107,ShafferPRX2021,RubtsovPRB2008,kotliar,pg3,pg6,pg9,HuangScience2019,BrownScience2019} and exact Bethe ansatz\cite{bethe} methods
affirm the presence of dispersing lower and upper Hubbard
bands in momentum space with a gap between them. Such
momentum space bands ultimately suggest that a momentum-space picture of the Mott problem must exist. Indeed, a formulation of the Mott problem along these lines would then put the standard formulation of metals and Mottness on the same footing, thereby solving the Mott problem.

It is precisely such a formulation that we address here. Hatsugai and Kohmoto\cite{hk} took the first step along these lines with their momentum-space analogue,
\begin{align}
    H_{\text{HK}} = \sum_{{\bf k},\sigma}\epsilon({\bf k})c^{\dagger}_{{\bf k}, \sigma}c_{{\bf k}, \sigma} + \sum_{\bf k}U n_{\bf k \uparrow}n_{\bf k \downarrow}
\end{align}
of the Hubbard model. This model is exactly solvable and yields an insulator anytime the interaction exceeds the bandwidth.  While this model lacks the momentum mixing of Hubbard, a continuous deformation between the two exists as has been recently formulated\cite{OHK}.   The trick\cite{OHK} is to put in all the momentum mixing the HK model leaves out.  We have shown that the momentum-mixing HK (MMHK) model\cite{OHK} leads to accurate results with minimal computational cost. 

What remains to be done is a formal mathematical proof that MMHK and Hubbard model have the same behavior in the charge sector.  We provide that proof here.

In constructing this proof, it is useful to review the MMHK\cite{OHK} scheme. For simplicity, we start with 2D band HK model. MMHK starts by introducing momentum scattering with the largest momentum transfer possible: $\bf k + (\pi,\pi)$.
The interaction that this introduces into the HK model is of the form,
\begin{align}
    H^{\text{MMHK}}_{\text{int},n=2}=\frac{U}{2}\sum_{ {\bf k} \in \text{BZ}} \sum_{{\bf P},{\bf Q} =(0,0),(\pi,\pi)}   c^{\dagger}_{ \bf k, \uparrow}c_{ \bf k - \bf Q ,\uparrow}c^{\dagger}_{ \bf k + \bf P , \downarrow}c_{ \bf k + \bf P + \bf Q , \downarrow},
\end{align}
Consider now the transformation: $c_{{\bf k}A\sigma}=\frac{1}{\sqrt{2}}(c_{{\bf k}, \sigma}+c_{{\bf k}+(\pi,\pi),\sigma})$ and $c_{{\bf k}B\sigma}=\frac{1}{\sqrt{2}}(c_{{\bf k}, \sigma}-c_{{\bf k}+(\pi,\pi),\sigma})$. In this new basis, $H^{\text{MMHK}}_{\text{int},n=2}$ is diagonal in $n_{{\bf k}A\sigma}$ and $n_{{\bf k}B\sigma}$, taking on the form,
\beq 
H^{\text{MMHK}}_{\text{int},N=2}&=U\sum_{{\bf k} \in \text{rBZ}_2}\sum_{a=\text{A},\text{B}}n_{{\bf k}a,\uparrow}n_{{\bf k}a,\downarrow}.
\eeq
However, under the new basis, the kinetic term becomes,
\beq
T^{\text{MMHK}}_{N=2} &= \sum_{{\bf k} \in \text{rBZ}_2 ,\sigma}\sum_{a,b = \text{A},\text{B}}[\epsilon({\bf k})]_{ab}c^{\dagger}_{{\bf k}a,\sigma}c_{{\bf k}b,\sigma} 
\eeq
where,
\beq
\epsilon(k)_{ab} & \equiv 
\frac{1}{2}[\epsilon({\bf k}) + (-1)^{\delta_{ab}}\epsilon({\bf k + (\pi,\pi) })]
\eeq
thereby breaking the commutativity enjoyed by the equivalent terms in the HK model.  It is from this non-commutativity that Hubbard dynamics arises.
The procedure of turning on the largest momentum transfer and downfolding the Brillouin zone can be done iteratively. After $n$ iterations, the resultant Hamiltonian,
\beq
    H^{\text{MMHK}}_{\text{int},n}=&\frac{U}{n}\sum_{{\bf k} \in \text{BZ}}\sum_{{\bf P},{\bf Q} \in \bf{B_n}} c^{\dagger}_{ \bf k, \uparrow}c_{ \bf k - \bf Q ,\uparrow}c^{\dagger}_{ \bf k + \bf P , \downarrow}c_{ \bf k + \bf P + \bf Q , \downarrow} ,
\eeq
contains an interaction term in which $2^n$ momenta are coupled and possible momentum transfer is restricted to a finite set $B_n$ with cardinality $\lvert B_n \rvert=n$. The reason why such a controlled expansion is achievable in ${\bf k}$ space instead of real space is that the HK model is controlled by a fixed point that maximally breaks the $Z_2$ symmetry\cite{ppfixedp} of a Fermi liquid.  

 Similarly, we find that $H^{\text{MMHK}}_{\text{int},n}$ is diagonal in the basis where $n$ momenta are grouped into a single cell.  To do this, we parametrize the whole Brillouin zone via $B_n$, the set of all possible momentum transfers. Then, any ${\bf k} \in BZ $ can be decomposed as a sum of ${\bf K} \in B_n$ and ${\bf q} \in rBZ_n \simeq BZ/B_n$.  Defining $\bf k = K + q$, and construct the new basis $c_{\bf qR \sigma}$,
\beq
    c_{\bf qR, \sigma}= \frac{1}{\sqrt{n}}\sum_{\bf K \in B_n}e^{i {\bf K \cdot R} }c_{\bf K+q,\sigma} \label{eq:FourieronK},
\eeq
through the appropriate Fourier transform on $\bf K$.
After a partial Fourier transform, we find that $H^{\text{MMHK}}_{\text{int},n}$ 
\beq
    H^{\text{MMHK}}_{\text{int},n}=U\sum_{{\bf q} \in \text{rBZ}_n}\sum_{\bf R}n_{{\bf qR},\uparrow}n_{{\bf qR},\downarrow},
\eeq
which lays plain\cite{ppfixedp} the diagonal structure and hence the $Z_2$-fixed point breaking structure.

Intuitively, when $n \to \infty$, we recover all momentum mixings and $\text{rBZ}_n$ shrinks to a point.  Hence, one intuitively expects that $\lim_{n\to \infty}H^{\text{MMHK}}_{\text{int},n}$ is related to the local in real space Hubbard interaction. Here, we present a rigorous proof that the $n \to \infty$ limit of n-MMHK is the Hubbard model with twisted boundary conditions.

To prove this requires an analysis of the kinetic part,
\beq
    T^{\text{MMHK}}_{n} =& \sum_{{\bf q} \in \text{rBZ}_n} \sum_{\sigma} \sum_{{\bf R_a,R_b}}[T({\bf q})]_{ab}c^{\dagger}_{{\bf q R_a},\sigma}c_{{\bf q R_b},\sigma},
    \eeq
    where, 
    \beq
    \left [T({\bf q})\right ]_{ab} \equiv& \frac{1}{n}\sum_{{\bf K \in B_n} }e^{-i \bf K \cdot (R_a - R_b)}\epsilon(\bf K+q).
\eeq 
The momenta are summed over ${\bf k} = (k_1,k_2,\dots ,k_d) \in {\text BZ}_n$, a $\text d$-dimensional vector in the Brillouin zone. For large enough $n$ and a smooth dispersion $\epsilon({\bf k})$, $|| {\bf q} || \le \frac{\pi}{n}$, always lies in the convergence radius of its Taylor series, which we write as
\beq
    \epsilon({\bf K+q}) = \sum_{n_1=0}^{+\infty}\sum_{n_2=0}^{+\infty} \dots \sum_{n_d = 0}^{+\infty}\frac{\epsilon^{(n_1,n_2,\dots ,n_d)}({\bf K})}{\prod _{m=1}^dn_m!}\prod_{m=1}^dq_m^{n_m},\nonumber\\
\eeq
where $\epsilon^{(n_1,n_2,\dots ,n_d)}({\bf K}) \equiv \partial_{K_1}^{n_1}\partial_{K_2}^{n_2}\dots \partial_{K_d}^{n_d}\epsilon ({\bf K}) = \prod _{m=1}^d \partial_{K_m}^{n_m}\epsilon ({\bf K})$.
We then perform the inverse transform of \eqref{eq:FourieronK},
\beq
    c_{{\bf K+q},\sigma}=\frac{1}{\sqrt{n}}\sum_{{\bf R}}e^{-i {\bf K \cdot R}}c_{{\bf q R},\sigma}.
\eeq
In this new basis, $ c_{{\bf qR},\sigma} $, the $(n_1,n_2,\dots ,n_d)$ order term  in the expansion is 
\beq
    \frac{1}{n}\sum_{{\bf K \in B_n}} \sum_{\bf R_1}\sum_{\bf R_2} \epsilon^{(n_1,n_2,\dots ,n_d)}({\bf K})e^{i {\bf K \cdot (R_1-R_2)}}c^{\dagger}_{{\bf qR_1},\sigma}c_{{\bf qR_2},\sigma}.\nonumber\\
\eeq

In the limit $n \to \infty$, the summation can be well approximated by an integration: $\frac{1}{n}\sum_{\bf K \in B_n} \simeq \int \frac{d^dK}{(2\pi)^d}$. This allows us to perform integration by part on $\epsilon^{(n_1,n_2,\dots ,n_d)}({\bf K})$, which yields 
\begin{widetext}
\beq
    \int \frac{d^dK}{(2\pi)^d}e^{i {\bf K \cdot (R_1-R_2)}}\prod_{m=1}^d \partial_{K_m}^{n_m}\epsilon ({\bf K}) 
    = \int \frac{d^d K}{(2\pi)^d}e^{i {\bf K \cdot (R_1-R_2)}}\prod _{m=1}^d [-i({\bf R_1-R_2})_m]^{n_m}\epsilon({\bf K}).
\eeq
\end{widetext}
We then perform the summation over $(n_1,n_2,\dots n_d)$,
\begin{widetext}
\beq
\sum_{n_1=0}^{+\infty}\sum_{n_2=0}^{+\infty}\dots \sum_{n_d = 0}^{+\infty} \prod _{m=1}^d\frac{[-iq_m({\bf R_1-R_2})_m]^{n_m}}{n_m!}=e^{-i {\bf q \cdot (R_1-R_2)}}
\eeq, 
\end{widetext}
which yields a phase factor. The exchange of integration and summation is allowed due to analyticity of the Taylor expansion. Thus, in the new basis $c_{{\bf nR},\sigma}$,
\beq
    \int\frac{d^dK}{(2\pi)^d}\sum_{\bf R_1,R_2}\epsilon({\bf K})e^{i {\bf (K-q) \cdot (R_1-R_2)}}c^{\dagger}_{{\bf qR_1},\sigma}c_{{\bf qR_2},\sigma}.
\eeq
Recall that for a tight-binding model with n sites and hopping amplitude $t_{\bf R_1,R_2}$, 
\beq
    \epsilon({\bf K})=\frac{1}{n}\sum_{\bf R_1,R_2}e^{-i{\bf K \cdot (R_1-R_2)}}t_{\bf R_1,R_2},
\eeq
the integration over K yields $t_{\bf R_1,R_2}$. We  conclude then that within the new basis, $c_{{\bf qR},\sigma}$, 
\beq
    T_{{\bf q},\sigma}=\sum_{\bf R_1,R_2}t_{\bf R_1,R_2}e^{-i {\bf q \cdot (R_1-R_2)}}c^{\dagger}_{{\bf qR_1},\sigma}c_{{\bf qR_2},\sigma}.
\eeq

Hence, we have shown that the $\text{nMMHK}$ model in the large $n$ limit is equivalent to a Hubbard model with a twist in the boundary conditions.  That is,
\beq
H_{\text{nMMHK}}=\sum_{{\bf q} \in \text{rBZ}_n} \sum_{\sigma}T_{{\bf q} \sigma}+H^{\text{MMHK}}_{\text{int},n}.
\eeq
One may read off the twist from the hopping phase $e^{-i {\bf q \cdot (R_1-R_2)}}$. For fixed $\bf q$ and lattice of size ${\bf N}=(N_1,N_2,\cdots, N_d)$, the corresponding twist in the boundary condition is $(q_1N_1,q_2N_2,\dots q_dN_d)$, since ${\bf q} \in \text{rBZ}_n$.

We now invoke the theorem by Watanabe\cite{Watanabe2014a} to see if any of the gapped structure which is present in HK and any momentum-mixing version survives the twist in the boundary conditions. He has shown that a twist in the boundary conditions cannot lead to a gap closing in the Kallen-Lehmann spectrum \cite{WatanabeTwist}. Although the original conclusion only applies to a gapped Hamiltonian, which is not the case for Hubbard model since its spin sector is gapless, one can still apply Watanabe's theorem\cite{WatanabeTwist} by restricting his inequality in Eq. (2) to the charge sector. The key to this proof is exponential fall-off of the correlations in the presence of a spectral gap.   As shown by Hastings\cite{hastings2006spectral}, operators in the charge sector exhibit exponentially decaying correlations as a result of the charge or Mott gap in the spectral function.  Consequently, if the $n \to \infty$ limit of $\text{MMHK}$ has a charge gap, then so does the Hubbard model.  This argument also works in reverse.
Since the MMHK construction applies to any dimension, we have shown that all Hubbard models possess a charge gap that smoothly interpolates between the weak and strong-coupling limits at half-filling.

The error incurred in converting sums to integrals can be estimated explicitly. In fact, the error between $\text {nMMHK}$ and $\text n$-site Hubbard has two origins. One is from the replacement of the summation, $\frac{1}{n}\sum_{\bf K \in B_n}e^{i \bf K \cdot R}f(\bf K)$ with the integral, $ \int \frac{d^dK}{(2\pi)^d}e^{i \bf K \cdot R}f(\bf K)$.  The second is from the twist in the boundary conditions.

We first address the replacement of the summation with an integral.  Since all the integrands are defined on the Brillouin zone, any Fourier transform of the form,
\begin{align}
    f({\bf K})=\sum_{{\bf l}}e^{i{\bf K \cdot l}}c_{{\bf l}},
\end{align}
is allowed.  Here ${\bf l} = (l_1,l_2,\dots,l_d)$ labels each Fourier component and $c_{\bf{-l}}=c^{*}_{{\bf l}}$,with a physical interpretation of $c_{\bf l}$ being the hopping amplitude from site $\bf r$ to site $\bf r+l$.
We may calculate $\Delta I_n({\bf R})$, the error between the discrete sum $\frac{1}{n}\sum_{\bf K \in B_n}e^{i\bf K \cdot R}f(\bf K)$ and the continuous integral $\int \frac{d^d \bf K}{(2 \pi)^d}e^{i \bf K \cdot R}f(\bf K)$. Consider this difference for each Fourier component $e^{i \bf K \cdot l}$,
\begin{widetext}
\beq 
\Delta I_{\bf n; l}(\bf R)=\frac{1}{n}\sum_{\bf K}e^{i\bf K \cdot (R + l)} - \int \frac{d^dK}{(2 \pi)^d}e^{i\bf K \cdot (R + l)} = \sum_{\bf Q \in \otimes_{i=1}^d N_i Z} \delta_{ \bf R+l,Q}-\delta_{\bf R+l}.
\eeq
\end{widetext}
We then calculate $\Delta I_n(\bf R)$ by summing over the error of each Fourier component, $\Delta I_n(\bf R)=\sum_{{\bf l}}\Delta I_{\bf n;l}({\bf R})  c_{l}$,
\beq
    \Delta I_n({\bf R})&=\sum_{\bf Q \in \otimes_{i=1}^d N_iZ }\delta_{\bf R+n, Q}c_{\bf n}-c_{-\bf R}.
\eeq
For a model with finite-range hopping $\bf n_B$, $c_{\bf n} = 0$ for $\bf n > n_B$.
Consequently, these two terms exhibit no difference $ \forall {\bf i},\bf N_i > n_B$. Hence, the error due to twisted boundary conditions is  exponentially small in the $\bf N \to \infty $ limit \cite{WatanabeTwist}.

We have proved that the recently formulated MMHK model differs from Hubbard by just a twist in the boundary conditions.  As shown by Watanabe\cite{Watanabe2014a}, any two gapped models that share only this difference must have the same bulk properties.  As MMHK shows a charge gap so does Hubbard. Consequently, MMHK is an accurate simulator of the charge sector of Hubbard physics.

\textbf{Acknowledgements:} We thank Erica Macgee and Amir Ibrahim for characteristically level-headed remarks on the convergence of MMHK to Hubbard.

\bibliographystyle{unsrt}
\bibliography{mottbib}

\end{document}